\documentclass[11pt]{article}
\usepackage{float}
\usepackage{array}
\usepackage{cool}
\usepackage{bbold}
\usepackage{eulervm}
\usepackage{hyperref}
\usepackage{tabularx}
\hypersetup{colorlinks=true,linkcolor=blue,citecolor=blue}
\let\oldmb\mathbold
\protected\def\mathbold{\oldmb}
\usepackage{amsmath}
\usepackage{todonotes}
\usepackage{graphicx}
\usepackage{subfig}
\usepackage{color}
 \usepackage{multicol}
\begin{document}
\title{\bf Magneto-tunable terahertz absorption in single-layer graphene: A general approach }

\author{D. Jahani \footnote{\href{d.jahani@sharif.edu}{d.jahani@sharif.edu}}, O. Akhavan, A. Alidoust Ghatar}
\maketitle {\it \centerline{
\emph{Department of Physics, Sharif University of Technology, P.O. Box 11155-9161, Tehran, Iran}}
\maketitle {\it \centerline{
\emph{ Leibniz Institute of Photonic Technology (IPHT), Jena, Germany. }}


\begin{abstract}
\emph{Terahertz (THz) anisotropic absorption in graphene could be significantly modified upon applying a static magnetic field on its ultra-fast 2D Dirac electrons. In general, by deriving the generalized Fresnel coefficients for monolayer graphene under applied magnetic field, relatively high anisotropic absorption for the incoming linearly polarized light with specific scattering angles could be achieved. We also prove that the light absorption of monolayer graphene corresponds well to its surface optical conductivity in the presence of a static magnetic field. Moreover, the temperature-dependent conductivity of graphene makes it possible to show that a step by step absorption feature would emerge at very low temperatures. We believe that these properties may be considered to be used in novel graphene-based THz application.}
\end{abstract}
\vspace{0.5cm} {\it \emph{Keywords}}: \emph{Graphene; Terahertz light; Optical absorption; Magnetic field.}
\section{ Introduction}
\emph{The electromagnetic waves technology, especially in THz region ranging from 0.1 to 10 THz is of most interest in many fields and applications particularly in remote control, sensing and detectors which are mostly related to absorption properties of the material [1,2,3,4,5]. Moreover, materials that strongly interact with light giving appropriate and desirable responses in a broad range of frequencies play a critical role in science and technology such as, photonics security, photodetectors, sensors, photovoltaics and absorbers [6,7]. Therefore, existence of an material with hight optical absorption in order to enhance absorption properties of structures could the most desirable aim in applications. Isolated in 2004, graphene has shown to be one of the ideal materials in THz light absorption application [8].}
\par
\emph{Graphene, an atomic layer of carbons, has recently attracted enormous interest due to its exceptional electronic and optical properties [9,10,11]. Many researchers believe that graphene can be an appropriate substitute candidate for silicon in the next high-speed generation of photonic and electronic components, owing to its high carrier mobility, tunable optical conductivity, and uniform light absorption over a wide-band wavelength [12,13]. Graphene can absorb incoming beams in the range of different wavelengths from visible to infrared, while conventional semiconductors are unable to absorb this range [14,15,16]. Bare graphene layer, owing to its unique band structure and massless carriers, approximately could absorb $2.3$ percent of light [17,18,19]. Note that this amount of absorption is so impressive for a material with a $0.34\ nm$ thickness. However, in general this amount of light absorption is considered to be very low and far from being used for desirable goals over a broad spectrum especially at the far-infrared and THz spectral ranges [17,20,21].}
\par
\emph{So far, several techniques have been introduced in order to enhance greatly the amount of light absorption of graphene in terms of theoretical and experimental approaches [22,23]. A long list of methods can be funded in at [20,21] for one that is interested to study which every method has its advantages and applications. Some methods are based on placing or embedding graphene in photonic crystals in order to achieve higher absorption [22,23,24,25]. Moreover, the optical absorption of graphene has been proved to be possible in a nanocavity resonator [26, 27]. Also perfect graphene absorber consisting of dielectric multilayer structures based on prism coupling  has been reported [28,29,30]. Furthermore, it has been shown that, tuned by varying the Fermi level, a graphene-based absorber could be realized by a periodic double-layer graphene ribbon structure in infrared region [31]. However, no research proved the relativley high absorbtion of bare graphene by applying a static magnetic field in the quantum regime for which optical interband transitions and Landau levels (LLs) play a central role in light interaction with its massless Dirac electrons. Here, thanks again to the optical and electrical properties of graphene which can be tuned by external factors, we demonstrate that it is accessible to efficiently improve the absorption performance of bare graphene layer in the qunatum Hall regime [32]. }
\par
\emph{In this study, we aim to achieve multi- and broadband high light absorption in a bare graphene layer by applying a constant magnetic field since it could be so sufficient to be used in many optical applications. This paper is organized as follows: we discuss the model and calculating in section 2, then numerical calculations and results are addressed in section 3. In the end, we summarize our findings in section 4.}

\section{Structure and theoretical modeling}
\emph{In this section, we provide a brief description of our calculation for incident $s$ and $p$ polarized light on a 2D conducting surface sandwiched between two dielectric media. In particular, we developed the components of the electric and magnetic vector of the incident, transmitted, and reflected waves as below [33]:
\begin{eqnarray}
\left\{\begin{array} {cc}
E_{i}=(-a_{p}\cos\theta_{i}, a_{s}, a_{p}\sin\theta_{i})e^{i\tau_{i}}\\
H_{i}Z_{0}=(-a_{s}n_{1}\cos\theta_{i}, -a_{p}n_{1}, a_{s}n_{1}\sin\theta_{i})e^{i\tau_{i}} \end {array} \right.
\end{eqnarray}
\begin{eqnarray}
\left\{\begin{array} {cc}
E_{r}=(-r_{p}\cos\theta_{r},r_{s}, r_{p}\sin\theta_{r})e^{i\tau_{r}}\\
H_{r}Z_{0}=(-r_{s}n_{1}\cos\theta_{r}, -r_{p}n_{1}e^{i\tau_{r}}, r_{s}n_{1}\sin\theta_{r})e^{i\tau_{r}} \end {array} \right.
  \end{eqnarray}
\begin{eqnarray}
\left\{\begin{array} {cc}
E_{t}=(-t_{p}\cos\theta_{t}, t_{s}e^{i\tau_{t}}, t_{p}\sin\theta_{t})e^{i\tau_{t}}\\
H_{t}Z_{0}=(-t_{s}n_{2}\cos\theta_{t}, -t_{p}n_{2}e^{i\tau_{t}}, t_{s}n_{2}\sin\theta_{t})e^{i\tau_{t}} \end {array} \right.
  \end{eqnarray}
Here, $a$, $r$, and $t$ are the complex amplitudes of the incidence, reflected, and transmitted waves with $\tau=\omega t-k.r$. Then boundary conditions relate the electric and magnetic field components at the interface, $z=0$ which can be described by following expressions::
\begin{eqnarray}
\left\{\begin{array} {cc}
E_{x}^{t}=E_{x}^{i}+E_{x}^{r} \ \ \ \ \ \ \ \ \ \ E_{y}^{t}=E_{y}^{i}+E_{y}^{r}, \\
H_{x}^{t}=H_{x}^{i}+H_{x}^{r}+J_{y}, \ \ \ \ \ \ \ \ \ \ H_{y}^{t}=H_{y}^{i}+H_{y}^{r}-J_{x} \end {array} \right.
  \end{eqnarray}
where $J=\bar\sigma .E^{t}$. Here, $J$ and $\sigma$ are the surface current density and the optical conductivity of 2D conducting material which we consider in to be graphene under an applied external magnetic field for which one can write the conductivity tensor as:
\begin{eqnarray}
\sigma=
\begin{pmatrix}
 \sigma_{xx} & \sigma_{xy} \\
 \sigma_{yx} & \sigma_{yy}
\end{pmatrix}
=
\begin{pmatrix}
 \sigma_{0} & \sigma_{H} \\
 -\sigma_{H} & \sigma_{0}
\end{pmatrix}
\end{eqnarray}
where $\sigma_{0}$ and $\sigma_{H}$, illustrate the longitudinal and Hall conductivity, respectively.
By considering equations (2.1), (2.2), (2.3), (2.4) and the relation $cos\theta_{r}=-\cos\theta_{t}$, reflection and transmission coefficients could be written in the following form:
\begin{eqnarray}
\left\{\begin{array} {cc} r_{ss}=(-1+\frac{2n_{1}G_{1}\cos\theta_{i}}{G_{1}G_{2}-Z_{0}^{2}\sigma_{yx}\sigma_{xy}\cos\theta_{i}\cos\theta_{t}})a_{s},\\
r_{sp}=(\frac{2n_{1}Z_{0}\sigma_{yx}\cos\theta_{i}\cos\theta_{t}}{G_{1}G_{2}-Z_{0}^{2}\sigma_{yx}\sigma_{xy}\cos\theta_{i}\cos\theta_{t}})a_{p},\\
r_{pp}=(1-\frac{2n_{1}G_{2}\cos\theta_{t}}{G_{1}G_{2}-Z_{0}^{2}\sigma_{xy}\sigma_{yx}\cos\theta_{i}\cos\theta_{t}})a_{p},\\
r_{ps}=(-\frac{2n_{1}Z_{0}\sigma_{xy}\cos\theta_{i}\cos\theta_{t}}{G_{1}G_{2}-Z_{0}^{2}\sigma_{xy}\sigma_{yx}\cos\theta_{i}\cos\theta_{t}})a_{s} \end {array} \right.
\end{eqnarray}
\begin{eqnarray}
\left\{\begin{array} {cc} t_{ss}=(\frac{2n_{1}G_{1}\cos\theta_{i}}{G_{1}G_{2}-Z_{0}^{2}\sigma_{yx}\sigma_{xy}\cos\theta_{i}\cos\theta_{t}})a_{s},\\
t_{sp}=(\frac{2n_{1}Z_{0}\sigma_{yx}\cos\theta_{i}\cos\theta_{t}}{G_{1}G_{2}-Z_{0}^{2}\sigma_{yx}\sigma_{xy}\cos\theta_{i}\cos\theta_{t}})a_{p},\\
t_{pp}=(\frac{2n_{1}G_{2}\cos\theta_{i}}{G_{1}G_{2}-Z_{0}^{2}\sigma_{xy}\sigma_{yx}\cos\theta_{i}\cos\theta_{t}})a_{p},\\
t_{ps}=(\frac{2n_{1}Z_{0}\sigma_{xy}\cos^{2}\theta_{i}}{G_{1}G_{2}-Z_{0}^{2}\sigma_{xy}\sigma_{yx}\cos\theta_{i}\cos\theta_{t}})a_{s} \end {array} \right.
\end{eqnarray}
where $G_{1}$ and $G_{2}$ are defined as $G_{1}=n_{1}\cos\theta_{2}+n_{2}\cos\theta_{1}+Z_{0}\sigma_{xx}\cos\theta_{1}\cos\theta_{2}$ \ and\ $G_{2}=n_{1}\cos\theta_{1}+n_{2}\cos\theta_{2}+Z_{0}\sigma_{yy}$. Here, $Z_{0}\approx 377 \Omega $ \ is the vacuum impedance and $a_{s}$\ ($a_{p}$) is the amplitude of the electric vector of the incident field in the perpendicular (parallel) plane, respectively. Note that the subscripts of pp, ps and ss, sp indicate that
incoming beams are linearly p- and s-polarized, respectively. However, pp, ps and ss, sp represent that the transmitted and reflect waves are p-, s- and s-, p- polarized, respectively. Now, as we use the quantum model for describing the conductivity, the longitudinal and Hall parts of the optical conductivity tensor ($\sigma$) of Dirac fermions with effective velocity $V_{F}$ at the temperature $T$ are as follows [34]:
\begin{equation}\begin{split}
\sigma_{0}(\omega)=
&\frac{e^{2}v_{f}^{2}\vert eB\vert\left(\hbar\omega+2i\Gamma \right)}{\pi i}\times\\& \sum_{n=0}^{\infty}\left\lbrace\frac{\left[f_{d}(M_{n})-f_{d}(M_{n+1})\right]+ \left[f_{d}(-M_{n+1})-f_{d}(-M_{n})\right]}{\left(M_{n+1}-M_{n}\right)^{3}-\left(\hbar\omega+2i\Gamma \right)^{2}\left(M_{n+1}-M_{n}\right)}\right\rbrace \\&
+ \left\lbrace\frac{\left[f_{d}(-M_{n})-f_{d}(M_{n+1})\right]+ \left[f_{d}(-M_{n+1})-f_{d}(M_{n})\right]}{\left(M_{n+1}+M_{n}\right)^{3}-\left(\hbar\omega+2i\Gamma \right)^{2}\left(M_{n+1}+M_{n}\right)}\right\rbrace
 \end{split}\end{equation}
 and:
 \begin{equation}\begin{split}
&\sigma_{H}(\omega)=\frac{-e^{2}v_{f}^{2} eB}{\pi}\\&
\sum_{n=0}^{\infty}\left\lbrace \left[f_{d}(M_{n})-f_{d}(M_{n+1})\right]- \left[f_{d}(-M_{n+1})-f_{d}(-M_{n})\right]\right\rbrace \times \\&
\left\lbrace \frac{1}{\left(M_{n+1}-M_{n}\right)^{2}-\left(\hbar\omega+2i\Gamma \right)^{2}}+ \frac{1}{\left(M_{n+1}+M_{n}\right)^{2}-\left(\hbar\omega+2i\Gamma \right)^{2}}\right\rbrace
\end{split} \end{equation}
in which $M_{n}=\sqrt{2n\vert eB\vert\hbar v_{f}^{2}}$. The scattering rate $\Gamma $ is assumed to be independent of the light frequency and LLs index and the distribution function is $f_{d}(M_{n}=1/(1+exp[(M_{n}-\mu)/K_{B}T])$ with $e$, $\hbar$ and $K_{B}$ representing electron charge and reduced Planck constant, Boltzmann constant, respectively. Finally, by the use of the above equations, the absorbtion of graphene for the incoming $s/p$-polarized light in quantum Hall regime could be given by:
\begin{eqnarray}
A_{s/p}=1-(Rss/pp+Rps/sp+Tss/pp+Tps/sp)
\end{eqnarray}
}
\section{Results and discussion}
\emph{Our absorber structure is just based on a single-layer graphene under quantum Hall effect situation. However, as a matter of more illustration we also consider graphene layer to be on a dielectric substrate as it is demonstrated schematically in Fig.1. Here, the dielectric substrate is considered to be $SiC$ (with the dielectric constant $4.4$) on which graphene could be epitaxially grown with a controlled number of atomic layer []. First, in our simulations, we consider the incoming TE polarization (s-polarized) to obtain the absorption performance of graphene in a relatively low magnetic field for the proposed structure. We show the results in an energy interval ranging from $10\ meV$ to $140\ meV$ for the chemical potential $4\ meV$ and Faraday geometry in which the external magnetic field is perpendicular to surface of graphene, i.e. along the propagation of incoming linearly polarized light (the case $\theta=0$). It is clear from Fig.2 that graphene in the magnetic fields reveals a multi-band absorption performance. When the magnetic field is set to be $B=1\ T$ and $B=3\ T$, the absorption of a graphene layer in the presence of a substrate has the amount of about $22 \%$ and $28.5 \%$ corresponding to the $76.53\ meV$ and $132.6\ meV$ photon energy, respectively. However, for bare graphene layer the absorption ratio is $42\%$ and $49\%$ for $B=1\ T$ and $B=3\ T$, respectively. Now, if we take attention to the effective optical conductivity of graphene for the same parameters and $B=1\ T$, we see that boosting in absorption is directly related to the nature of the optical conductivity of graphene. In Fig. 3, We show the real and imaginary parts of the effective optical conductivity of graphene. As it is seen, each absorption peaks (for graphene with and without a substrate) in the energy position, exactly corresponding to the peaks of the optical conductivity of graphene. Consequently, this result indicates that the absorption performance of bare graphene could be tuned by modifying its optical conductivity. }
\par
\emph{The absorption of single-layer graphene under an increasing applying magnetic field for different angles of the incoming light for $\mu=0.4\ eV$ and $E_{ph}=20\ meV$ at a very low temperature $T=1\ K$ and the room temperature $T=300\ K$ is depicted in Fig. 4. We, in Fig. 4 (a), observe a step-like scheme emerges for the absorption of graphene as a function of the magnetic field ranging from $B=1\ T$ to $B=50\ T$. It is clear that by increasing the magnetic field the absorption is also increased until it reaches its ultimate amount. Then this trend is reversed and the absorption falls gradually by increasing the field. In Fig. 4 (b) the role of scattering angle in the absorbtion is more clear. Note that, as it is clear from Fig. 4 (c) at a higher temperature the absorption will be boosted to some extend and the step-like scheme tends to vanish. The situation is more illustrated in Fig.4 (d).}
\par
\emph{In the following, the effect of the chemical potential ranging from $\mu=0.1\ eV$ to $\mu=0.8\ eV$ on the absorption performance on graphene has been surveyed. The outcomes are illustrated as a function of the angle of the incidence of incoming linearly polarized light and also two different temperatures. For graphene with a substrate ( $E_{ph}=20\ meV$ and $B=20\ T$) similar to the magnetic field case, a plateau-like shape for the absorption can be seen (Fig. 5 (a)). The absorption is increased step by step until about $\mu=0.471\ eV$, then it begins to fall. Note that, in spite of the fact that the highest plateau has happened in the chemical potential between $0.471\ eV$ and $0.511\ eV$, the maximum absorption is occurred at $\mu=0.491\ eV$ with $19.38\%$ at $\theta=0$. From Fig. 5(c) it is clear that enlarging angle of the incoming light leads to the lower absorption for graphene. Besides, at high temperatures, the absorption rate is boosted up to $21.19 \%$ and the step-like structure, as it might be expected, also tends to be faded. In Fig. 6, the absorbtion versus the chemical potential for graphene without a substrate in a similar condition as indicated in Fig. 5, have been examined. In comparison to Fig. 5, the step-like situation is also taking place and the absorption is increased by increasing the chemical potential. }
\par
\emph{At this point, to see how the photon energy ($E_{ph}$) ranging from $0.1\ meV$ to $70\ meV$ affects the absorption of graphene, we also examine it as a function of the incident angles at $B=20\ T$ and $\mu=0.4\ eV$ for $T=1\ K$ and $T=300\ K$. For graphene layer as it is observed from Fig. 7(a), the absorption of incoming s-polarized light shows a peak at a specific angle. Therefore, we see that $A_{s}$ increases when the direction of the applied magnetic field makes an angle to the propagation of the incident light relative to the Faraday geometry for a specific light's scattering angle. However, the absorption of the incoming p-polarized incident light increasing by increasing the scattering angle leads to lower values for $A_{p}$ (see Fig. 8).   }
\section{Conclusion}
\emph{In summery, we investigated the absorbtion spectrum of intraband and interband transitions through the surface conductivity tensor of graphene in its general anisotropic state. We have numerically detailed the optical absorption of single layer graphene in THz region with and without a substrate ($SiC$) under quantum Hall effect situation. Since, in general, the initial polarization of light will no be preserved under the presence of the magnetic field, we have examined the effect of increasing the static magnetic field, chemical potential and photon energy interval for different incident angles of the incoming linearly polarized light. It was shown that any peak in the absorption performance of graphene layer in the photon energy interval directly corresponds to the enhancement of the optical conductivity of monolayer graphene. Therefore, we see that enhancement of the absorbtion of bare graphene is directly related to the profile of the optical conductivity of graphene. However, one can use photonic resonates and other resonance methods to increase the absorbtion of graphene-based devices. Significantly, applying a magnetic field on garpehene results in appearing a step-like scheme in the absorbtion of graphene at low temperatures. Moreover, we proved that higher absorbtion for bare graphene can be achieved for special incident angles of incident light relative to the Faraday geometry which may open opportunities for further investigation of light absorbtion in graphene-based THz application. }
\section{Data availability}
Data sharing is not applicable to this article as no new data were created or analyzed in this study.

\begin{figure}
\begin{center}
\includegraphics[width=18cm]{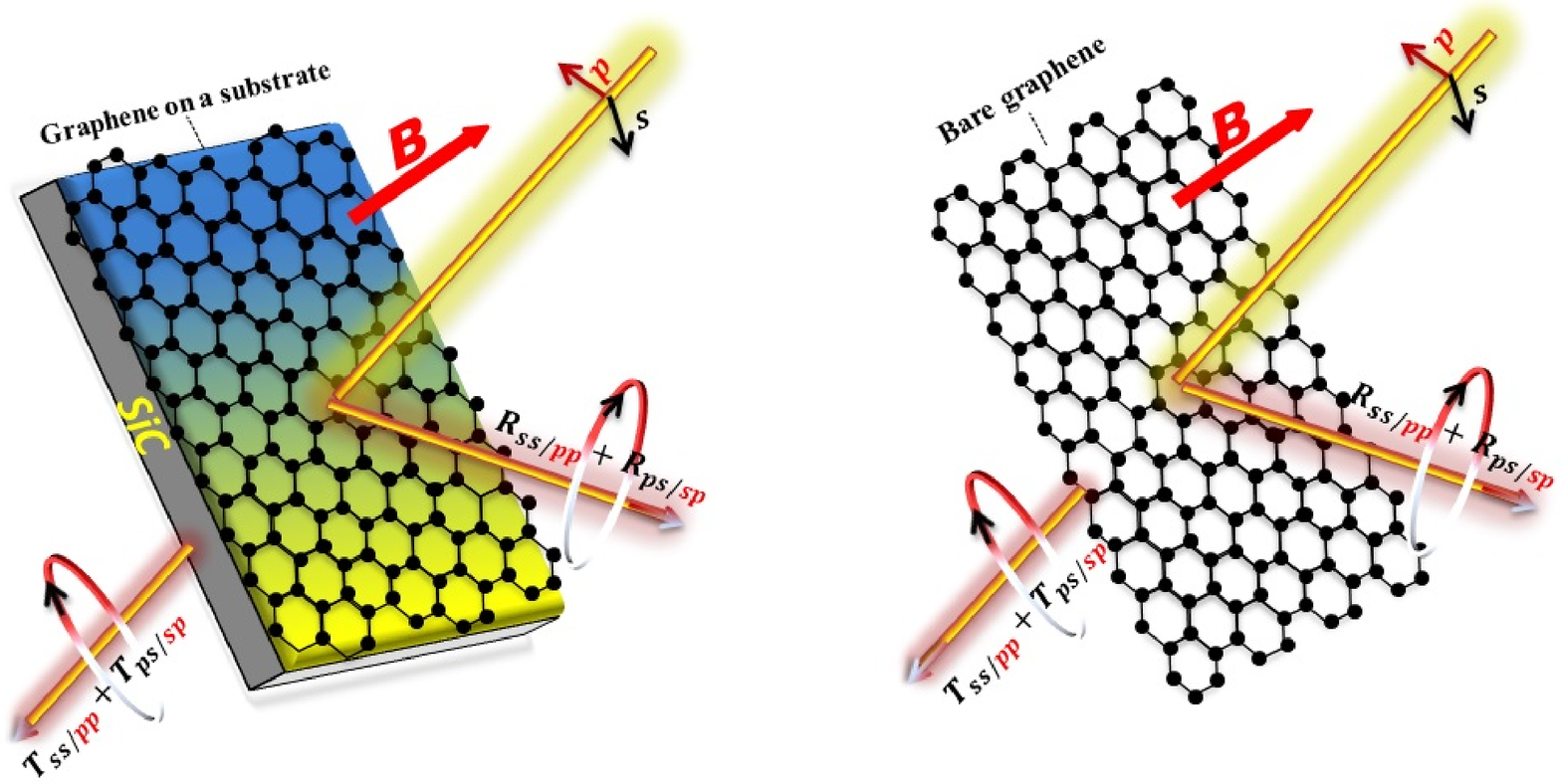}
 \caption{Schematic representation of light's incidence on graphene in a general non-Faraday geometry (the perpendicular external magnetic field applied on the surface of graphene is not along the propagation of the incident light).}

\end{center}
\end{figure}

\begin{figure}
\begin{center}
\includegraphics[width=18cm]{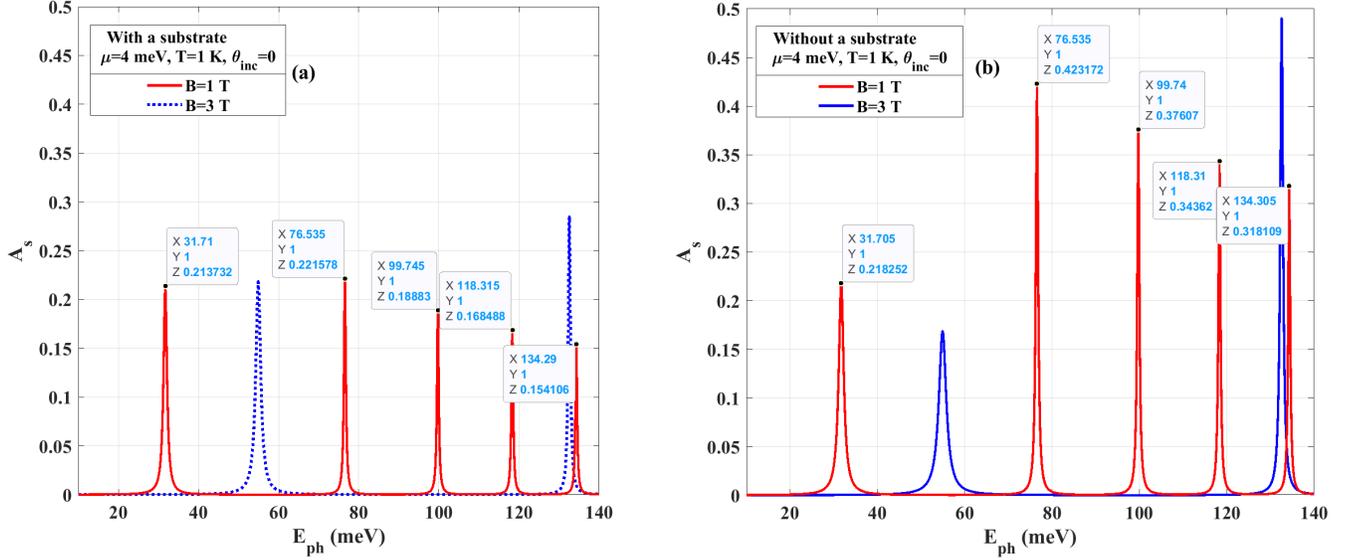}
 \caption{(a) Absorbtion of bare graphene verses the photon energy interval ranging from $0.1\ meV$ to $140\ meV$ for $B=1\ T$ (red curve) and $B=3\ T$ (Blue curve) at $T=1\ K$ for the incident light. (b) Absorbtion of graphene on $SiC$ verses the photon energy. We see that absorbtion for bare graphene could marks higher values than the case when it is put on a substrate.}
\end{center}
\end{figure}

\begin{figure}
\begin{center}
\includegraphics[width=18cm]{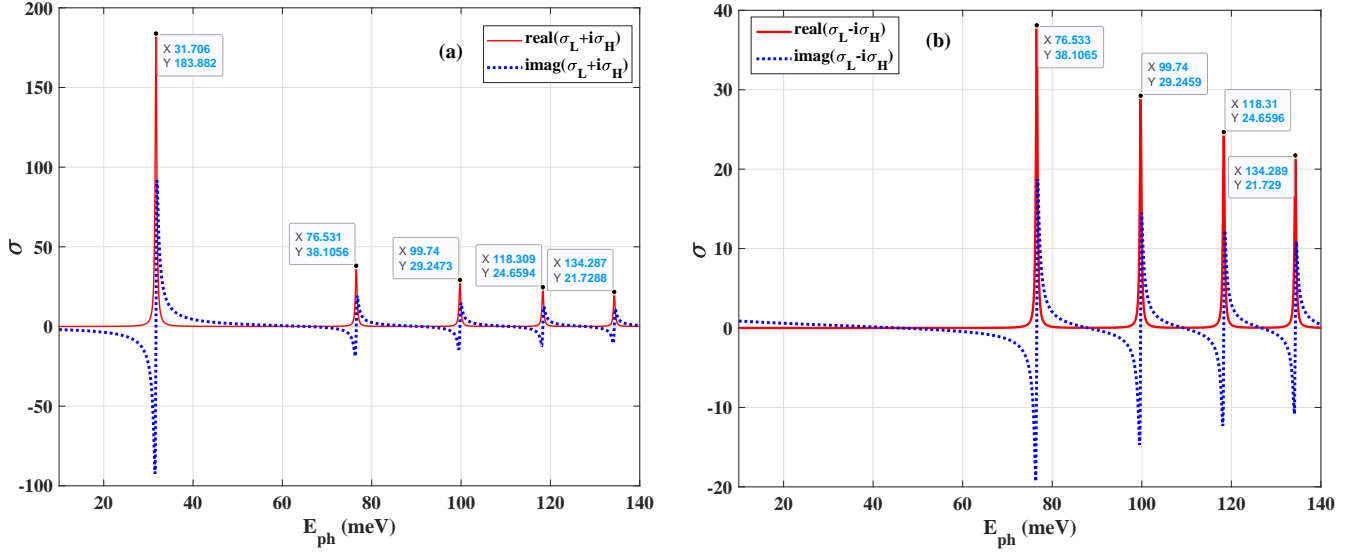}
 \caption{(a) and (b) The real and imaginary part of effective conductivity as a function of the photon energy $E_{ph}$ with the parameters used as in Fig.2 for left and right handed waves. As it is clear, the peaks in the optical conductivity exactly correspond to the peaks seen in the absorption.}
\end{center}
\end{figure}

\begin{figure}
\begin{center}
\includegraphics[width=18cm]{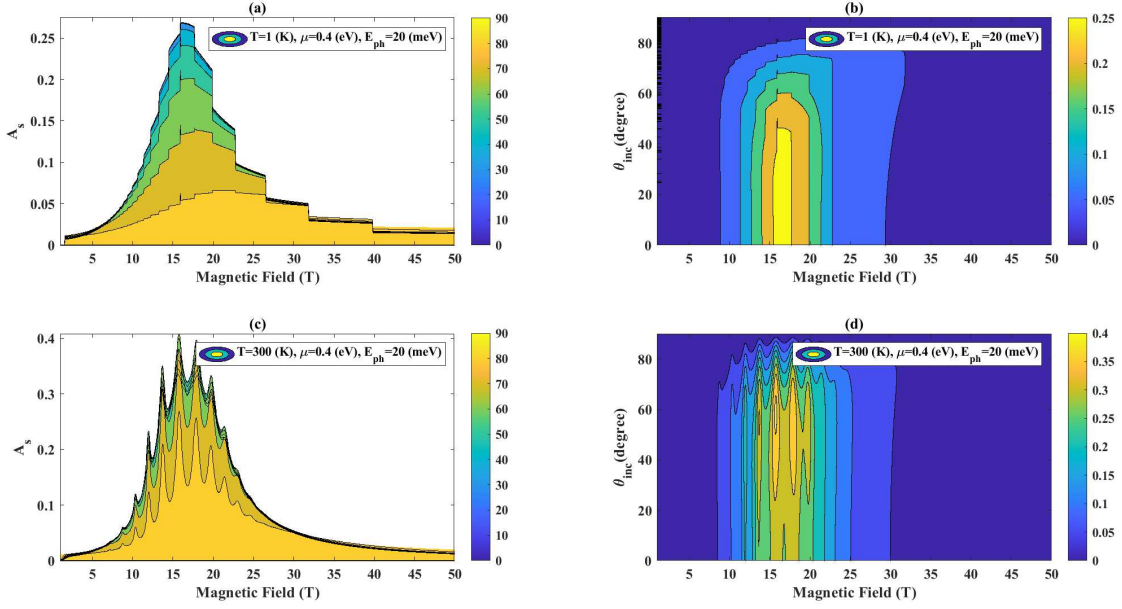}
\caption{(a) and (b) Absorption spectra under the effect of increasing magnetic field for different incident angles of incoming $s$-polarized THz light at $T=1\ K$. (c) and (d) Absorption at the high temperature ($T=300\ K$) for graphene. As it is expected, a step-like scheme for the absorption appears at low temperatures. However, when the temperature increases the step-like feature vanishes. The absorbtion for higher incident angles shows to be increased for s-polarized light.}
\end{center}
\end{figure}

\begin{figure}
\begin{center}
\includegraphics[width=18cm]{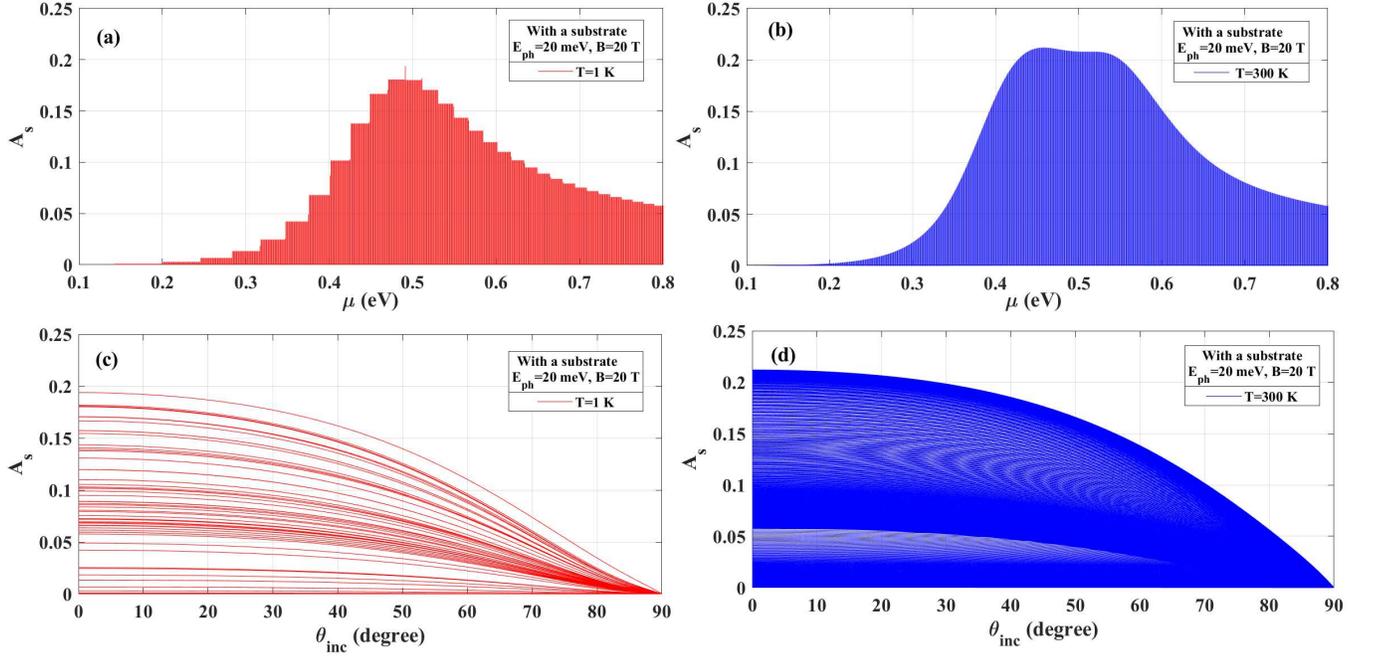}
 \caption{Variation in the absorption versus the chemical potential at $T=1\ K$ (a) and $T=300\ K$ (b) for graphene on $SiC$. Absorbtion as a function of the incident angle at $T=1\ K$ (c) and $T=300\ K$ (d) for graphene with a substrate for $E_{ph}=20\ meV$ and $B=20\ T$.}
\end{center}
\end{figure}

\begin{figure}
\begin{center}
\includegraphics[width=18cm]{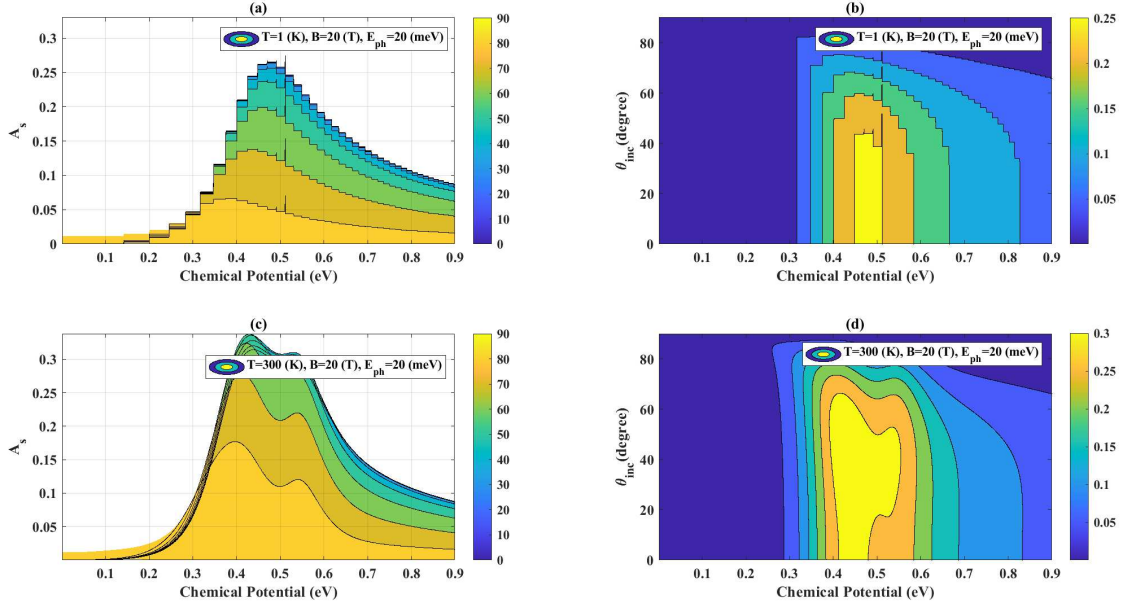}
\caption{(a) and (b) Absorption of bare graphene as a function of the chemical potential for $T=1\ K$. (c) and (d) Impact of incoming light for different scattering angles at $T=300\ K$ for bare graphene under the incident $s$-polarized light with $E_{ph}=20\ meV$ and $B=20\ T$.}
\end{center}
\end{figure}

\begin{figure}
\begin{center}
\includegraphics[width=18cm]{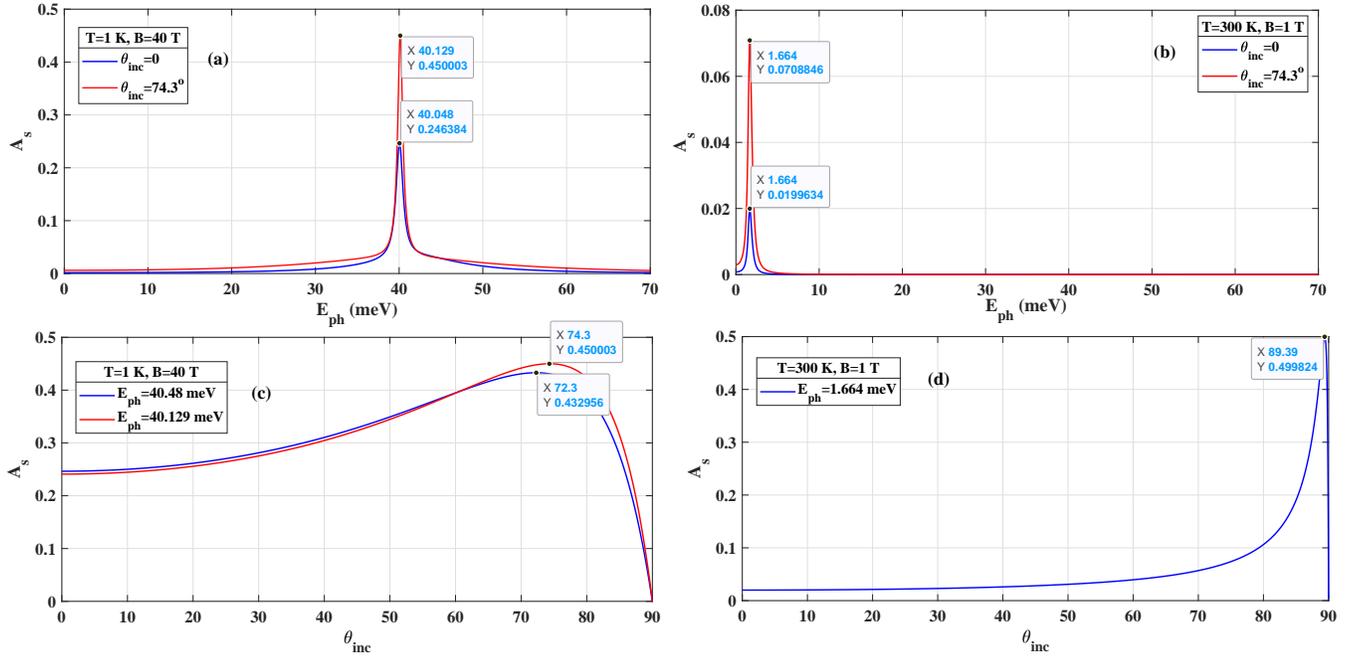}
\caption{(a) and (b) Absorption for two different scattering angle at $T=1\ K$ and $T=300\ K$, respectively. Absorption for different scattering angles at $T=1\ K$ (c) and $T=300\ K$ (d) for bare graphene for two different values for photon energy with $s$ polarization.}
\end{center}
\end{figure}
\begin{figure}
\begin{center}
\includegraphics[width=18cm]{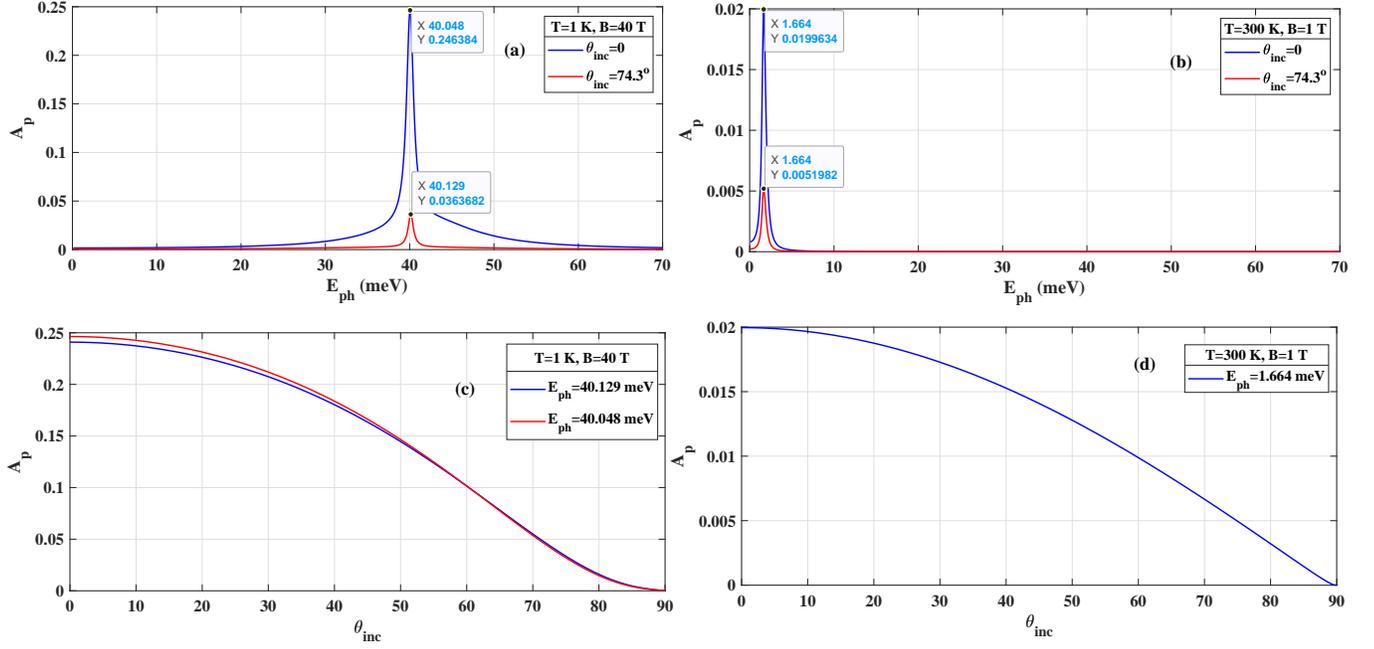}
\caption{(a) and (b) Absorption of incoming $p$-polarized light at $T=1\ K$ and $T=300\ K$, respectively. Impact of incoming light in different angles at $T=1\ K$ (c) and $T=300\ K$ (d) for bare graphene for two different values for photon energy with $s$ polarization .}
\end{center}
\end{figure}


\end{document}